\documentclass{aastex}
\usepackage{amssymb,amsmath}
\def\app#1#2{%
	\mathrel{%
		\setbox0=\hbox{$#1\approx$}%
		\setbox2=\hbox{%
			\rlap{\hbox{$#1\propto$}}%
			\lower1.1\ht0\box0%
			}%
			\raise0.25\ht2\box2%
			}%
			}

\usepackage[hyphens]{url}
\usepackage{revsymb}
\begin{document}
%\sloppy
\title{Quasars {\it vs.\/} Microquasars: Scaling and Particle Acceleration}
\shorttitle{Quasars {\it vs.\/} Microquasars}
\shortauthors{Katz}
\author{J. I. Katz\altaffilmark{}}
\affil{Department of Physics and McDonnell Center for the Space Sciences,
Washington University, St. Louis, Mo. 63130}
\email{katz@wuphys.wustl.edu}
%\shortauthors{Katz}
%\date{\today}
%\hypersetup{draft}  %Permits references to break across columns
%\begin{document} %MNRAS
%\psfrag{theta}{$\theta$}  Works only on COMPLETE strings
%\label{firstpage} %MNRAS
%\pagerange{\pageref{firstpage}--\pageref{lastpage}} %MNRAS
%\maketitle %MNRAS
\begin{abstract}
Quasars and microquasars both contain accreting black holes and power
	nonthermal double radio sources, but differ in more than their
	scales: Quasars are proportionally much more efficient accelerators
	of energetic electrons.  The radio luminosity of {the double
	radio sources associated with quasars, reflecting the long-time
	average of the particle acceleration power, is often} 1--30\% of the
	{quasar's} bolometric luminosity; in microquasars the fraction
	is $\lesssim 10^{-5}$.  This may be explained by {the scaling of
	accretion disc parameters with the black hole mass}.
\end{abstract}
\newpage
\section{Introduction}
The first quasars recognized were 3C~48 and 3C~273 \citep{S63}.  \citet{S64}
explained their enormous luminosities as the result of accretion onto
supermassive black holes.  Quasars are typically accompanied by twin opposed
radio lobes, often with narrow jets along the paths of energetic electrons
accelerated near the black hole that power the synchrotron emission of the
extended radio lobes.  Particle acceleration is ubiquitous in low density
astronomical plasmas, but no predictive theory exists \citep{K91}.

Micro-quasars are scaled-down analogues of quasars with black holes of
stellar mass.  The type specimens are 1E1740.7$-$2942 \citep{M92} and GRS
1915$+$105 \citep{MR94,T04,CHATY}.  Despite their qualitative similarities
(variability, nonthermal emission and double synchrotron radio lobes),
quasars and microquasars differ quantitatively: quasars convert a much
larger fraction of their bolometric luminosity to particle acceleration.  I
summarize the data for a few exemplars and offer a qualitative explanation
derived from accretion disc scaling laws.  \citet{LB08} found a related but
distinct correlation between the radio and X-ray luminosities of
a number of classes of astronomical objects.
\section{Data}
{A very large number of quasars exist because most massive galaxies
appear to have supermassive black holes at their centers, and accretion onto
these black holes makes a quasar.  In contrast, recent summaries
\citep{CHATY,microlist} list only about 19 microquasars.  The available data
are incomplete for most objects in both categories.  Table~\ref{data} shows
some of the most famous and best-studied quasars and those microquasars for
which bolometric and radio luminosities (the result of acceleration of
relativistic electrons) are known.}

The Table shows the ratio of radio luminosity, an estimate of the power that
accelerates energetic electrons {averaged over their (long) lifetime in
the radio lobes}, to bolometric luminosity, taken as the X-ray luminosity of
microquasars and the visible luminosity of quasars, for exemplars of each
type (considering Sco X-1 as a microquasar; \citep{F83}).  Both categories
are heterogeneous and show a wide range of values, and the sources are
individually very variable, but these data show that quasars convert orders
of magnitude greater fractions of their bolometric luminosity to
accelerating relativistic electrons than do microquasars.
\subsection{Sco X-1}
\label{scox-1}
The Galactic binary X-ray source Sco X-1 was subsequently discovered
\citep{F83} to be accompanied by radio lobes like those of quasars, though
on energetic and geometric scales many orders of magnitude smaller.  The
compact object in Sco X-1 is argued, on the basis of its mass inferred from
light curve modeling \citep{CKB22}, to be a neutron star rather than a black
hole.

Yet no evidence of a spin period in Sco X-1 has been reported, as would be
expected if it is a neutron star.  This may be explained either if the mass
of the compact object has been underestimated and it is, in fact, a black
hole, or if it is a neutron star with a magnetic field too weak to modulate
its accretion.  Radio pulsars are known with magnetic fields as small as
$\sim 2 \times 10^7$ G \citep{ATNF}.  This would be insufficient to disrupt
the accretion flow of $\sim 10^{18}$ g/s implied by the $\sim 10^{38}$ erg/s
X-ray luminosity of Sco X-1.  Even if uniformly distributed over a neutron
star's surface, the hydrodynamic stress of this accretion rate would be
$\sim 10^{15}$ dyne/cm$^2$, requiring a magnetic field $\gtrsim 10^8$ G to
channel the flow.

If Sco X-1 is such a weakly magnetized neutron star, perhaps a late stage in
the evolution of a recycled pulsar, its accretion disc and emission would be
expected to resemble, at least qualitatively, those of an accreting black
hole.  Whether a weakly magnetized neutron star or a black hole, we include
Sco X-1 in our list of micro-quasars; exclusion would not change the
conclusions of this paper.

\begin{table*}
       \begin{center}
        \begin{tabular}{|ccc|ccccc|}
%                \hline
                \multicolumn{3}{c}{Quasars}& \quad &\multicolumn{4}{c}{Microquasars}\\
%               \hline
                \cline{1-3}
                \cline{4-8}
		Cyg~A & 3C~273 & 3C~48 & \quad & Sco~X-1 & GRS 1915$+$105 & 1E1740.7$-$2942 & Mean\\
		0.3 & 0.015 & 0.03 & \quad & $10^{-9}$ & $2 \times 10^{-6}$ & $1.5 \times 10^{-8}$ & $9 \times 10^{-6}$ \\
                \hline
        \end{tabular}
        \end{center}
	\caption{\label{data} Fractions of bolometric luminosity that
	accelerates energetic electrons in exemplar objects.  Bolometric
	luminosity is visible light luminosity for quasars, X- and
	gamma-rays for microquasars.  ``Mean'' is the geometric mean
	fraction for four microquasars (LSI~$+61^\circ$303, LS~5039, SS~433
	and Cyg~X-3) in \citet{CHATY} with L-band and Integral 3--10 KeV
	data; individual values range from $2 \times 10^{-6}$ to $6 \times
	10^{-5}$.  Radio luminosities are in L-band except GRS 1915$+$105 in
	X-band.  Other data are from \citet{CB96,G08} (Cyg~A), \citet{T99}
	(3C~273), \citet{MS63,NRAO} (3C~48), \citet{F83} (Sco~X-1),
	\citet{T04} (GRS~1915$+$105), \citet{S91} and \citet{M92}
	(1E1740.7$-$2942).}
\end{table*}

\section{Condition for Particle Acceleration}
In order to accelerate energetic particles, necessary both for incoherent
emission (as in AGN and radio sources) and for coherent emission (as in
pulsars and fast radio bursts (FRB), if these particles drive a plasma
instability), it is necessary that they gain energy from an electric field
faster than they lose it to interaction with an ambient medium by ``Coulomb
drag'' \citep{A09}.  For a relativistic electron the ratio of these
quantities defines an acceleration parameter
\begin{equation}
        \label{A}
        A \approx {E m_e c^2 \over 4 \pi e^3 n_e \ln{\Lambda}},
\end{equation}
where $E$ is the electric field, $n_e$ the medium's electron density and the
Coulomb logarithm $\Lambda = 2 m_e c^2/I$, where $I$ is the ionization
potential in a neutral medium or $m_ec^2/\hbar \omega_p$ in a plasma.  In
most astronomical environments $\ln{\Lambda} \approx 20$.  $A > 1$ is
necessary for particle acceleration.
\section{Accretion Discs}
In the midplane of an accretion disc, a simple scaling model may be used to
estimate the scaling of the magnetic induction $B$.  The accretion rate
\begin{equation}
	\label{dotM}
	{\dot M} = 2 \pi h \rho r v_r,
\end{equation}
where $h$ is a disc thickness, $\rho$ the mass density and $v_r$ a radial
velocity.  The shear stress
\begin{equation}
	\sigma = {B^2 \over 8\pi}
\end{equation}
and the torque it exerts
\begin{equation}
	{\cal T} = {B^2 \over 8\pi} 2\pi r^2 h.
\end{equation}

The torque must remove the specific angular momentum $\ell = \sqrt{GMr}$ of
the accreting matter:
\begin{equation}
	\label{T}
	{\cal T} = {\dot M}\ell = \sqrt{GMr}{\dot M},
\end{equation}
where $M$ is the black hole mass.  The accretion rate $\dot M$ is related to
the bolometric luminosity $L = \epsilon {\dot M} c^2$ by the radiative
efficiency $\epsilon$, a dimensionless quantity plausibly independent of
$M$.  Combining the two expressions for the torque,
\begin{equation}
	\label{B}
	{B^2 \over 8 \pi} = {{\dot M} \over 2\pi} \sqrt{GM \over r^3}
	{1 \over h}.
\end{equation}
A characteristic value of the electric field in the near-vacuum above the
accretion disc is
\begin{equation}
	\label{E}
	E = {v \over c}B = \sqrt{GM \over rc^2} B =
	\sqrt{{GM \over r c^2}{4 {\dot M} \sqrt{GMr} \over r^2 h}}.
\end{equation}
%\begin{equation}
%        \label{B}
%        \begin{split}
%                {B^2 \over 8 \pi} &\sim {1 \over 2} {L \over L_E}
%                \left({GM \over r c^2} \right)^{3/2} {r \over h}
%                {c^4 \over GM \kappa} \\ &\sim 3 \times 10^7\ {L \over L_E}
%                \left({10 GM \over r c^2}\right)^{3/2} {r \over 10 h}
%                {10^8 M_\odot \over M}\ {{\rm erg} \over {\rm cm}^3},
%        \end{split}
%\end{equation}
%where it is assumed that the dissipative stress driving the viscous energy
%loss and emission of power $L$ is comparable to the magnetic stress
%$B^2/8\pi$; $r$ is the distance from the central mass $M$ and $h$ the local
%disc thickness.

In order to calculate $A$ it is necessary to estimate the electron density
$n_e$.  The component of gravity normal to the disc midplane is
\begin{equation}
	g = {GM \over r^2}{h \over r},
\end{equation}
so the disc pressure is approximated
\begin{equation}
	\label{P}
	P = {GM \over r^3} \rho h^2.
\end{equation}
Then the fractional disc thickness
\begin{equation}
	\label{thick}
	{h \over r} = \sqrt{{P \over \rho}{r \over GM}}.
\end{equation}
The shear stress
\begin{equation}
	\label{sigma}
	{B^2 \over 8\pi} = \alpha P = \alpha {h^2 \over r^3}{GM\rho}
	= {{\dot M}\ell \over 2 \pi r^2 h},
\end{equation}
where $\alpha$ is the ratio of shear stress to pressure.

Combining Eqs.~\ref{dotM} and \ref{sigma}, the radial velocity
\begin{equation}
	v_r = \alpha {h^2 \over r^2} \sqrt{GM \over r}.
\end{equation}
Using the right hand equality of Eq.~\ref{sigma},
\begin{equation}
	\rho = {{\dot M} \ell r \over 2 \pi h^3 GM \alpha}
\end{equation}
and the electron density 
\begin{equation}
	\label{n_e}
	n_e = {\rho \over \mu m_p} = {{\dot M} \ell r \over
	2 \pi h^3 GM \alpha \mu m_p} \propto M^{-1},
\end{equation}
where $m_p$ is the proton mass and $\mu$ the molecular weight per electron.
Then
\begin{equation}
	\label{result}
	A = {m_e c^2 \alpha m_p \mu \over e^3 \ln{\Lambda}}
	\sqrt{\epsilon \kappa \over 4 \pi}
	\left({GM \over r c^2}\right)^{1/4} \left({h \over r}\right)^{5/2}
	\sqrt{L_E \over L} \sqrt{GM}.
\end{equation}
The opacity $\kappa$ appears because the luminosity $L$ is scaled to the
Eddington luminosity $L_E \equiv 4 \pi cGM/\kappa$.  %For a fully ionized

The preceding relations are not quantitative but may give useful information
about how their properties scale.  The first factors in Eq.~\ref{result} are
either fundamental constants or dimensionless parameters that may be
independent of $M$.  The final factor gives the scaling of $A$:
\begin{equation}
        \label{Ascale}
        A \propto M^{1/2}.
\end{equation}
More massive black holes provide a more favorable environment for the
acceleration of energetic electrons.

Some of the results of this section were presented, in different forms, by
\citet{K06}
\section{Discussion}
%This paper only presents scaling relations and
%the variables are characteristic quantities, defined only qualitatively.
%The justification for this is that the microphysics of accretion discs is
%not well understood, and that the goal of this paper is to explain
%differences of several orders of magnitude in the efficiency of particle
%acceleration between systems whose black holes also differ in mass by
%several orders of magnitude.
The central mystery of quasars is why so much of their power appears as
particle acceleration, energizing their double radio lobes, rather than
being radiated thermally by the accretion disc.   If we did not know that
that accretion onto supermassive black holes makes powerful nonthermal radio
sources, we might expect them to be thermal emitters.  This cannot explain
acceleration of relativistic particles {or the existence of their double
radio lobes}.  The low density corona or funnel of the disc around a
supermassive black hole, too dilute for magnetohydrodynamics to be
applicable, offers a favorable environment for the acceleration of energetic
particles.  Eq.~\ref{Ascale} explains why supermassive black hole accretion
is a more favorable environment for particle acceleration than accretion
onto stellar mass black holes, as indicated by the data in the Table.
The electron density $n_e$ must be orders of magnitude less than the
mid-plane estimate Eq.~\ref{n_e} used in deriving Eq.~\ref{result} in order
that $A \gg 1$.  Such small $n_e$ may be found in a gravitationally
stratified disc corona or centrifugally stratified disc funnel.

%This cannot be calculated quantitatively.  The scaling of Eq.~\ref{Ascale}
%may explain why quasars (AGN) are proportionally more efficient accelerators
%of energetic particles than microquasars and accretion discs around stellar
%mass black holes, as shown in the Table.
In denser plasma electric fields produce current flow that screens out
time-varying magnetic fields and reduces the electric field from the value
in Eq.~\ref{E}, but in very low density plasma above the accretion disc or
in its throat the current density is limited to $n c e$ and may be
insufficient to exclude the magnetic field, justifying Eq.~\ref{E} as a
scaling relation.

At lower plasma densities ``anomalous'' resistivity, resulting from
two-stream electron-positron or ion-acoustic instability, may be large,
preventing the cancellation of the induced electric field by current flow.
These instabilities accelerate energetic particles.  In the coron\ae\ or
axial throats of accretion discs, particularly those around supermassive
black holes in which the electron density is comparatively small, the
``anomalous'' resistivity may be high or the current saturated because of
the low density.  Then Eq.~\ref{E} may be, at least approximately, valid and
Eq.~\ref{result} useful for scaling.

If repeating FRB are produced by microquasars \citep{K22a} their disc
rotation axes are likely aligned with the line of sight.  The immediate
vicinities of their black holes will be persistent radio sources, as
observed in some, but not all, repeating FRB.  Unless the disc axes precess
\citep{K22b} or drift over time, their radio emission would not appear as
double lobed sources because the lobes would be aligned with the disc
rotation axes and our lines of sight.  The known persistent radio sources
associated with some repeating FRB are not spatially resolved, consistent
with this hypothesis.
%\section*{Compliance with Ethical Standards}
%The author has no potential conflicts of interest.
\section*{Data Availability}
This theoretical study generated no new data.

\end{document}